\documentclass[twocolumn,8pt]{article}

\hypersetup{breaklinks=true,colorlinks=true,citecolor = blue}

\usepackage[super,comma,sort&compress]{natbib}

\usepackage{amssymb}
\usepackage[fleqn]{amsmath}

\usepackage{times} 

\usepackage{graphicx}

\usepackage{xcolor}
\definecolor{shadecolor}{RGB}{84,148,158}

\usepackage[font=footnotesize,labelfont=bf]{caption}
\renewcommand{\refname}{\small References}

\newcommand{\apj}{Astrophys. J.}
\newcommand{\apjl}{Astrophys. J. Lett.}
\newcommand{\apjs}{Astrophys. J. Suppl. Ser.}
\newcommand{\aap}{Astron. Astrophys.}
\newcommand{\aj}{Astron. J.}
\newcommand{\mnras}{Mon. Not. R. Astron. Soc.}
\newcommand{\nat}{Nature}

\newcommand{\jcap}{J. Cosmol. Astropart. P.}
\newcommand{\prd}{Phys. Rev. D}
\newcommand{\procspie}{Proc. SPIE}

\begin{document}

\baselineskip=5mm

\small

\renewcommand{\figurename}{Fig.}

\twocolumn[

\noindent\colorbox{shadecolor}
{\parbox{\dimexpr\textwidth-2\fboxsep\relax}{\textsf{\large \color{white}{LETTERS}}}}\\
\smallskip

\textsf{\textbf{\Large Gravitational lensing detection of an extremely dense environment around a galaxy cluster}}\\ 

\noindent
\normalsize{\textsf{\textbf{Mauro Sereno$^{1,2,*}$, 
Carlo Giocoli$^{2,1,3}$, 
Luca Izzo$^4$, 
Federico Marulli$^{2,1,3}$, 
Alfonso Veropalumbo$^{2,3}$, 
Stefano Ettori$^{1,3}$, 
Lauro Moscardini$^{2,1,3}$, 
Giovanni Covone$^{5,6}$, 
Antonio Ferragamo$^{7,8}$, 
Rafael Barrena$^{7,8}$, 
Alina Streblyanska$^{7,8}$
}
}
}\\

\textsf{\footnotesize
$^1$INAF - Osservatorio di Astrofisica e Scienza dello Spazio di Bologna, via Piero Gobetti 93/3, I-40129 Bologna, Italia\\
$^2$Dipartimento di Fisica e Astronomia, Alma Mater Studiorum -- Universit\`a di Bologna, via Piero Gobetti 93/2, I-40129 Bologna, Italia\\
$^3$INFN, Sezione di Bologna, viale Berti Pichat 6/2, I-40127 Bologna, Italia\\
$^4$Instituto de Astrof{\`i}sica de Andaluc{\`i}a (IAA-CSIC), Glorieta de la Astronom{\`i}a s/n, E-18008 Granada, Spain\\
$^5$Dipartimento di Fisica, Universit\`{a} di Napoli `Federico II', Compl. Univers. di Monte S. Angelo, via Cinthia, I-80126 Napoli, Italia\\
$^6$INFN, Sezione di Napoli, Compl. Univers. di Monte S. Angelo, via Cinthia, I-80126 Napoli, Italia\\
$^7$Instituto de Astrof\'isica de Canarias (IAC), C/ V\'ia L\'actea s/n, E-38205 La Laguna, Tenerife, Spain\\
$^8$Universidad de La Laguna, Departamento de Astrof\'isica, C/ Astrof\'isico Francisco S\'anchez s/n, E-38206 La Laguna, Tenerife, Spain\\
$^*$e-mail: \href{mailto:mauro.sereno@oabo.inaf.it}{mauro.sereno@oabo.inaf.it}\\
}

]

\smallskip

\noindent
\textsf{\textbf{Galaxy clusters form at the highest density nodes of the cosmic web\cite{kai84,sh+to04}. The clustering of massive halos is enhanced relative to the general mass distribution and matter beyond the virial region is strongly correlated to the halo mass (halo bias)\cite{tin+al10}. Clustering can be further enhanced depending on halo properties other than mass (secondary bias)\cite{wec+al06,dal+al08,li+al08,mao+al17}. The questions of how much and why the regions surrounding rich clusters are over-dense are still unanswered\cite{joh+al07,ser+al15_bias,mor+al16,zu+al17,bu+wh17,dvo+al17}. Here, we report the analysis of the environment bias in a sample of very massive clusters, selected through the Sunyaev-Zel'dovich effect by the  {\it Planck} mission\cite{planck_2015_XXVII,ser+al17_psz2lens}. We present the first detection of the correlated dark matter associated to a single cluster, PSZ2~G099.86+58.45. The system is extremely rare in the current paradigm of structure formation. The gravitational lensing signal was traced up to 30 megaparsecs with high signal-to-noise ratio ($\sim3.4$). The measured shear is very large and points at environment matter density in notable excess of the cosmological mean. The boosting of the correlated dark matter density around high mass halos can be very effective. Together with ensemble studies of the large scale structure, lensing surveys can picture the surroundings of single haloes.
}
}

The environment bias $b_\text{e}$ expresses the matter overdensity in the surroundings of massive halos. Observational campaigns have ascertained the halo bias and its mass dependence\cite{joh+al07,ser+al15_bias,dvo+al17} but efforts to detect enhancing mechanisms or secondary biases for massive halos have been inconclusive. Contamination by foreground or background groups hampers the analysis in stacked cluster subsamples\cite{zu+al17,bu+wh17}. Here, we measure for the first time the environment bias of a single massive cluster by detection of the weak lensing (WL) signal. WL distorts the shape of the background galaxies. The correlated matter around the halo imprints a peculiar feature in the shear profile\cite{og+ta11}. No proxy is needed. Mass and concentration of the halo and environment bias can be determined by fitting the shear profile up to very large distances. Even though the measurement of the shear around a single cluster is very challenging due to high noise, the interpretation is much more direct than for stacked samples, where the noise is reduced at the price of averaging over heterogeneous or contaminated samples.

Sunyaev-Zel'dovich (SZ) selected galaxy clusters appear to be trustful tracers of the massive end of the cosmological halo mass function\cite{ros+al16}. We studied the environment bias in the PSZ2LenS sample\cite{ser+al17_psz2lens}, which consists of the 35 galaxy clusters detected by the {\it Planck} mission\cite{planck_2015_XXVII} in the sky portion covered by the lensing surveys CFHTLenS (Canada France Hawaii Telescope Lensing Survey)\cite{hey+al12} and RCSLenS (Red Cluster Sequence Lensing Survey) \cite{hil+al16}. PSZ2LenS is a statistically complete and homogeneous subsample of the PSZ2\cite{planck_2015_XXVII} catalogue. It is approximately mass limited and the main halo properties are in excellent agreement with the $\Lambda$CDM (Cold Dark Matter with a cosmological constant $\Lambda$) scenario of structure formation \cite{ser+al17_psz2lens}.

\begin{figure}
\resizebox{\hsize}{!}{\includegraphics{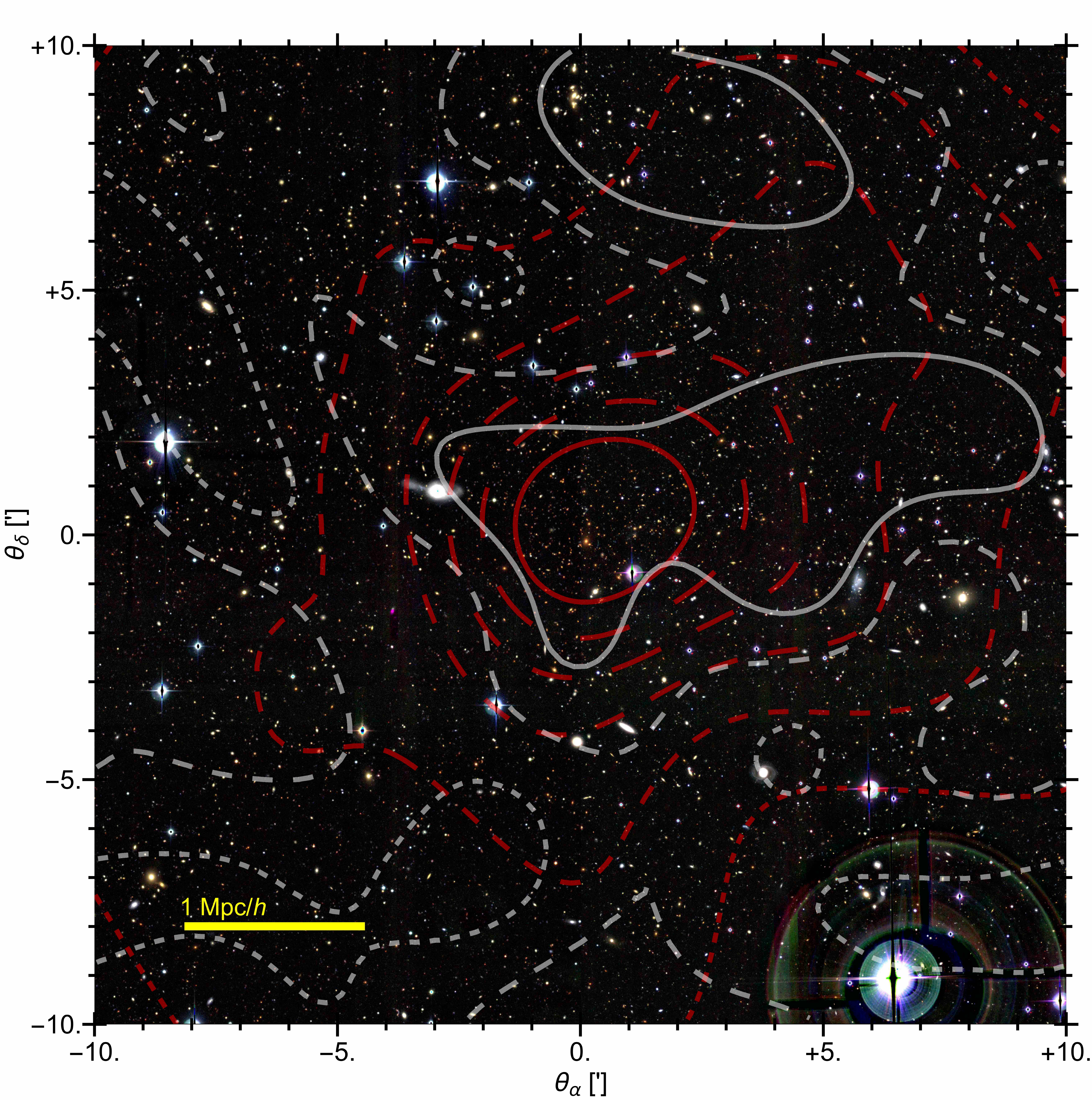}}
\caption{\textbf{Visible light and total mass}. Shown is the colour image of PSZ2~G099.86+58.45 and surroundings. The contours show the mass distribution reconstructed from WL (white), and optical $i$-light of the galaxies with photometric redshift within $\pm0.06(1+z_\text{cl})$ of the cluster redshift (red). The longer the dash, the higher the contour value. The map is centred on the BCG. North is up.}
\label{fig_PSZ0478_map}
\end{figure}

The WL quality data up to very large radii enables us to investigate the environment bias. Our main target is PSZ2~G099.86+58.45, the highest redshift cluster, $z_\text{cl}=0.616$, of the PSZ2LenS sample, see Fig.~\ref{fig_PSZ0478_map}. PSZ2~G099.86+58.45 is a hot cluster. The temperature of the intra-cluster medium is $T_\text{X}=8.9^{+2.8}_{-1.1}$~keV, as derived from the spectroscopic analysis of {\it XMM}-Newton data. The galaxy velocity dispersion is $\sigma_\text{v}=680^{+160}_{-130}\text{km~s}^{-1}$.

The cluster is located at the centre of the CFHT-Wide 3 field. The angular diameter distance to the cluster is very significant, $D_\text{d}\simeq 0.98~\text{Gpc}~h^{-1}$, where $h=H_0/(100~\text{km}~\text{s}^{-1}~\text{Mpc}^{-1})$, which is $\sim 80$\% of the maximum angular diameter distance reachable in our universe. The shear can then be investigated up to very large proper projected distances from the cluster centre, which we identify with the dominant brightest cluster galaxy (BCG). However, the lens redshift is still such that we can measure the shape distortion of a significant number of faint sources in the background. This makes PSZ2~G099.86+58.45 an ideal target.

\begin{figure}
\resizebox{\hsize}{!}{\includegraphics{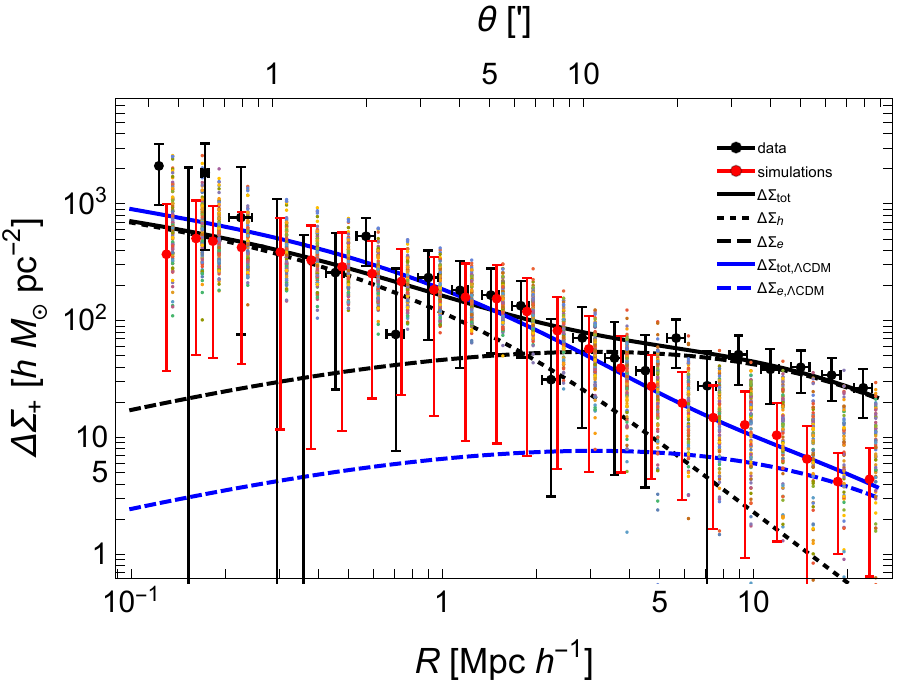}}
\caption{\textbf{Lensing profile.} Shown is the differential surface density $\Delta\Sigma_+$ of PSZ2~G099.86+58.45 as a function of the proper projected distance from the BCG. Black points are the measurements. The horizontal error bars are the weighted standard deviations of the distribution of radial distances in the annulus. The vertical error bars include statistical and LSS noise. Red points are the averaged simulations; small coloured points are for single realisations. The vertical red bar includes 68.3\% of the simulated profiles. Red and coloured points are horizontally shifted for visualization purposes. The black curve is the best fitting profile on the full radial range. The green and blue curves plot the contribution by the main and the correlated matter, respectively. The dashed curves plot the $\Lambda$CDM model.
}
\label{fig_PSZ0478_Delta_Sigma}
\end{figure}

We recovered the mass distribution around the cluster with the WL analysis of the differential surface density $\Delta\Sigma_+$, see Fig.~\ref{fig_PSZ0478_Delta_Sigma}.  The shear signal was collected from more than 150000 galaxies  up to $\sim 25.1~ \text{Mpc}~h^{-1}$ ($\sim 1.46$ degrees).  

Shape distortions of galaxies were measured in CFHT wide-field images in the $i$ optical band\cite{erb+al13,mil+al13}; photometric redshifts were estimated exploiting observations in the $u$, $g$, $r$, $i$ and $z$ bands\cite{hil+al12,ben+al13}. We selected background galaxies using their colours or their photometric redshifts. The effective redshift of the selected source galaxies is $z_\text{s} \sim 0.96$. The overall level of shear systematics due to calibration errors, fitting procedure, contamination by foreground or cluster member galaxies, photometric redshift uncertainties and intrinsic alignment is at the $\sim5\%$ level.

The mass distribution as inferred from WL can be compared with the light density of the galaxies at the cluster redshift, selected if their photometric redshift is within $|z - z_\text{cl}|\le 0.06\times(1+z_\text{cl})$. Notwithstanding the poor angular resolution of the WL analysis, the map comparison tentatively suggests that matter peaks coincide with galaxies overdensities, see Fig.~\ref{fig_PSZ0478_map}.

We measured the WL signal in circular annuli, see Fig.~\ref{fig_PSZ0478_Delta_Sigma}. All the matter along the line of sight contributes to the lensing phenomenon. We can identify three main agents: (i) the main lens, i.e. the collapsed and nearly virialized cluster, which is dominant at radii $\lesssim 3~\text{Mpc}$;  (ii) the correlated matter in the surroundings ($\gtrsim 3~\text{Mpc}$), comprising the satellite halos, the filamentary structure and the smoothly accreting matter\cite{col+al05,eck+al15}; (iii) the uncorrelated matter of the large scale structure (LSS) which fills the line-of-sight.

The measured signal in the range $10<R~[\text{Mpc}~h^{-1}]<25.1$, where the correlated matter is the dominant term, is $\Delta \Sigma_{+,\text{obs}}=32.1\pm4.5(\text{stat.})\pm8.1(\text{LSS})\pm1.6(\text{sys.})~M_\odot~h~\text{pc}^{-2}$. The signal-to-noise ratio is $\text{SNR}\simeq 3.4$. This provides a clear and model independent detection of the cluster surroundings. 

The $\Lambda$CDM paradigm makes strong predictions on clusters and surroundings in terms of mass and redshift of the main halo. The cluster can be modelled with a cuspy density profile\cite{nfw96} whose mass and concentration are correlated\cite{men+al14}. The mass profile is truncated at the splash-back radius beyond which the matter is still infalling \cite{die+al17}. The correlated matter is dominant beyond the splash-back radius. It can be expressed as a 2-halo term\cite{og+ta11}, where the halo bias is a function of the peak height\cite{tin+al10}. The LSS contribution acts as a noise whose amplitude is fixed by the effective projected power spectrum\cite{sch+al98b}. It is significant on very large scales.

We reconstructed the matter distribution and compared it with $\Lambda$CDM predictions. We used two parametric modellings. Firstly, we fitted the shear profile only in the region more sensitive to the main halo ($0.1<R<3~ \text{Mpc}~h^{-1}$) with an informative prior on the mass-concentration relation\cite{men+al14} and environment bias $b_\text{e}$ modelled as a 2-halo term\cite{tin+al10}. This is the $\Lambda$CDM model, which gives mass $M_{200} \sim (8.3\pm3.2)\times 10^{14} M_\odot h^{-1}$, concentration $c_{200} =3.4.\pm 0.9$, and $b_\text{e,$\Lambda$CDM}=11.1\pm2.5$. The uncertainty on the theoretical fitting function of the halo bias is $\sim6\%$, as estimated from the simulation to simulation scatter \cite{tin+al10}, even though simulations poorly cover the mass and redshift range around PSZ2~G099.86+58.45 and some extrapolation is needed.

The measured WL mass $M_{500} \sim (5.5\pm2.0)\times 10^{14} M_\odot h^{-1}$ is in good agreement with expectations based on multi-probe proxies: $M_{\text{X},500} \sim (5.1 \pm1.8)\times 10^{14} M_\odot h^{-1}$ based on $T_\text{X}$; $M_{\text{SZ},500} \sim (6.1 \pm0.8)\times 10^{14} M_\odot h^{-1}$ based on the integrated Compton parameter\cite{planck_2015_XXVII}; $M_{\sigma,500} \sim (1.2 \pm0.8)\times 10^{14} M_\odot h^{-1}$ based on the galaxy velocity dispersion. 

Secondly, we fitted the full profile ($0.1<R<25.1~ \text{Mpc}~h^{-1}$) with the bias as a free parameter. Now, $ b_\text{e}$ quantifies how much the total matter around the halo is overdense. We found  $b_\text{e}=78\pm11$. Priors on mass and concentration of the main halo affects very marginally this bias estimate. The multi-probe analysis confirms that the mass measurement of the cluster is solid and that the bias excess cannot be explained in terms of an under-estimated halo mass.

The measured signal in the range $10<R~[\text{Mpc}~h^{-1}]<25$ is much larger than the average $\Lambda$CDM prediction, $\Delta \Sigma_{+,\Lambda\text{CDM}}=5.6\pm1.5~M_\odot~h~\text{pc}^{-2}$, hinting to two possible, not exclusive causes: very overdense correlated matter boosted by formation mechanisms or projection effects from uncorrelated matter. 

To quantify the degree of discrepancy, we performed numerical simulations exploiting the Lagrangian perturbation theory, where the hierarchical formation of dark matter halos is realised from an initial density perturbation field\cite{mon+al13}. This method is very effective in covering the mass and redshift range we are interested in, which can be challenging for standard $N$-body simulations.

\begin{figure}
\resizebox{\hsize}{!}{\includegraphics{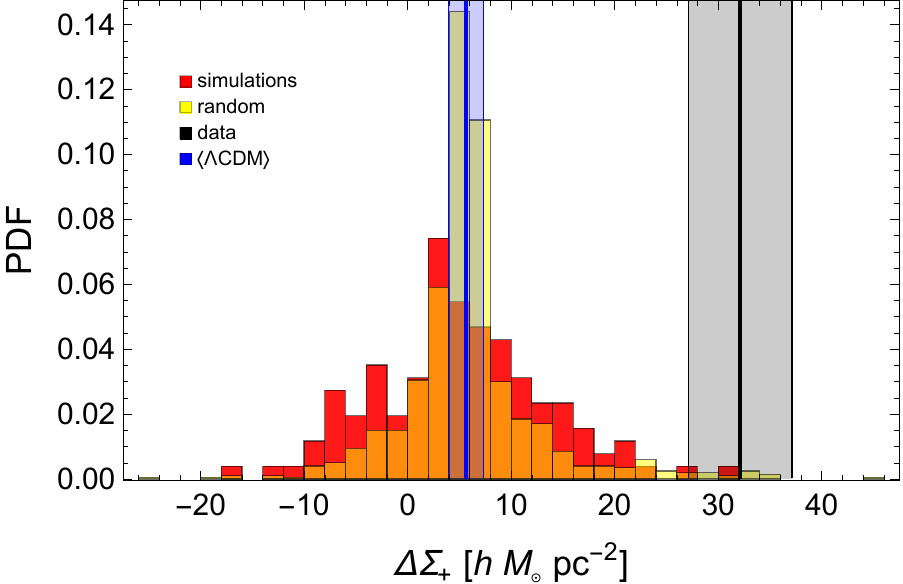}}
\caption{\textbf{Differential surface density of correlated matter around PSZ2~G099.86+58.45.}  Shown is $\Delta\Sigma_+$ in the radial range $10<R<25.1~ \text{Mpc}~h^{-1}$. The histograms show the theoretical predictions, as obtained from numerical simulations (red) or as the signal at random pointings added to the expected value (yellow). The black vertical lines mark the observed value for PSZ2~G099.86+58.45 (full black) and the 68.3\% confidence region (dashed). The blue line marks the average $\Lambda$CDM prediction.
}
\label{fig_PSZ0478_histo_DeltaSigma_sim}
\end{figure}

We derived the shear around 128 simulated halos in the redshift range $0.54\lesssim z \lesssim 0.71$ with average mass and redshift as PSZ2~G099.86+58.45. The signal is consistent with the analytical $\Lambda$CDM model, see Fig.~\ref{fig_PSZ0478_Delta_Sigma}. The probability that an overdense line of sight boosts the shear at the measured level is of $\lesssim 0.5\%$, see Fig.~\ref{fig_PSZ0478_histo_DeltaSigma_sim}. This is also confirmed by the analysis of the shear distortions at random locations in the CFHTLS-W3 field, where the measured signal is greater than the excess $\Delta \Sigma_{+,\text{obs}} - \Delta \Sigma_{+,\Lambda\text{CDM}}$ with a probability of only $\sim0.9\%$ (Fig.~\ref{fig_PSZ0478_histo_DeltaSigma_sim}).

The largest simulated shears are associated to overdense regions at the lens redshift. Whereas the typical simulated system shows multiple peaks of uncorrelated matter along the line of sight, the simulation with the highest signal, $\Delta \Sigma_{+,\text{sim}}\simeq31.0~M_\odot~h~\text{pc}^{-2}$, shows a single prominent peak at the lens redshift. Simulations strongly disfavour uncorrelated noise as the only source of signal excess, which is related to the cluster surroundings.

\begin{figure}
\resizebox{\hsize}{!}{\includegraphics{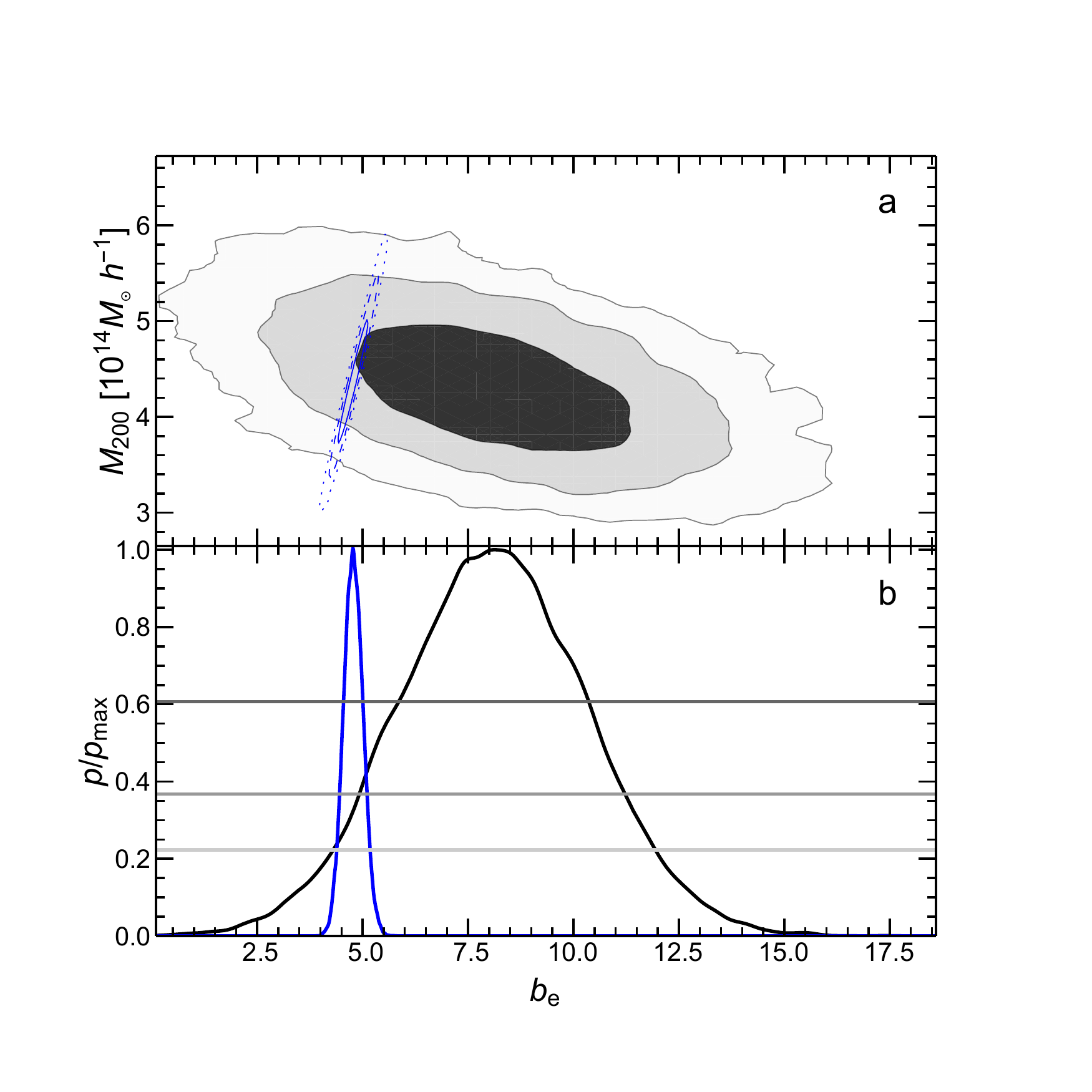}}
\caption{\textbf{Environment bias of PSZ2LenS.} Shown are the measured $b_\text{e}$ (black lines) and the $\Lambda$CDM prediction (red). The top panel shows the probability distribution in the bias-mass plane. The contours include the 68.3, 95.4 and 99.7\% confidence regions in two dimensions, here defined as the regions within which the probability is larger than $\exp[-2.3/2]$, or $\exp[-6.17/2]$ of the maximum, respectively. The bottom panel shows the marginalised one-dimensional distribution of the halo bias, renormalised to the maximum probability. The magenta, blue and green levels denote the confidence limits in one dimension, i.e. $\exp[-1/2]$, $\exp[-1]$ and $\exp[-2]$ of the maximum.
}
\label{fig_PSZ2LenS_bias_PDF}
\end{figure}

The shear excess can be also investigated for the full PSZ2LenS sample, see Fig.~\ref{fig_PSZ2LenS_bias_PDF}. We first stacked the shear measurements of the PSZ2LenS clusters, which lie either in the CFHTLenS or in the RCSLenS, and we then fitted the combined differential density profile\cite{ser+al17_psz2lens}. The effective lensing weighted redshift of PSZ2LenS is $z_\text{stack}=0.20$. As for PSZ2~G099.86+58.45, we fitted the signal with the $\Lambda$CDM modelling in the radial range $0.1<R~[\text{Mpc}~h^{-1}]<3.1$ or with $b_\text{e}$ as a free parameter for $0.1<R~[\text{Mpc}~h^{-1}]<25.1$.

The stacked analysis probes the bias at an unprecedented halo mass of $(4.3\pm0.5)\times10^{14}M_\odot h^{-1}$. Even though the stacked analysis cannot probe any assembly bias, since we combined the signal with no regard to the secondary halo properties, we can still look for bias excess in the favourable case of high SNR ($\sim3.9$) and low LSS noise, which is now greatly reduced by averaging over different sky regions. Even though the noise affecting the stacked sample is much smaller than for PSZ2~G099.86+58.45 ($1.1$ vs $9.3~M_\odot~h~\text{pc}^{-2}$), SNRs are comparable, which further stresses the extremely high signal produced by the surroundings of PSZ2~G099.86+58.45.

The $\Lambda$CDM expectation for the stacked PSZ2LenS sample ($b_\text{e,$\Lambda$CDM}=5.4\pm0.3$) is compatible with the measured environment bias of the stacked clusters ($b_\text{e}=8.1\pm2.2$) at the 11.9\% level, see Fig.~\ref{fig_PSZ2LenS_bias_PDF}, which is statistically significant. The result is not driven by PSZ2~G099.86+58.45, which has a low lensing weight given the high redshift and the relatively small number of background galaxies. In fact, the results do not change significantly if we exclude PSZ2~G099.86+58.45 from the stacking or we consider the subsample at low redshift.

The SZ selection is unique in sampling the massive end of the halo mass function and unveiling new cluster properties. The environment bias for the PSZ2LenS sample is statistically consistent with $\Lambda$CDM predictions and the correlated matter around PSZ2~G099.86+58.45 lies in the extreme value tail. This is an extremely rare case. Clustering around cluster-sized halos can be amplified for low halo concentrations or high spins or a significant number of subhalos with a large average distance, even tough it is still uncertain why and if these different proxies of halo assembly history can exhibit different trends\cite{mao+al17}. According to the statistics of peaks, the extreme environment bias of PSZ2~G099.86+58.45 is associated to a peak of the primordial Gaussian density field with a very low value of the curvature $s=|d\langle\delta\rangle/d\log M |$, where $\delta$ is the density fluctuation and $M$ is the mass\cite{dal+al08}.  These are the locations of larger background density and enhanced clustering for very massive halos. Formation and evolution mechanisms can be very effective in boosting the environment density. Next generation of galaxy surveys will routinely perform the lensing analysis of single halos out to very large radii that we have presented here for the first time.

\section*{\small Methods}

\footnotesize

\footnotesize

\noindent \textbf{GL signal.} Our analysis exploited the public CFHTLenS and RCSLenS shear catalogs. WL data were processed with \texttt{THELI}\cite{erb+al13} and shear measurements were obtained with \texttt{lensfit}\cite{mil+al13}. We computed the differential projected surface density $\Delta\Sigma_+$ in circular annuli as
\begin{equation} 
\label{eq_Delta_Sigma_3}
\Delta \Sigma_+ (R) = \frac{\sum_i  (w_i \Sigma_{\text{cr},i}^{-2}) \epsilon_{\text{+},i} \Sigma_{\text{cr},i}} {\sum_i ( w_i \Sigma_{\text{cr},i}^{-2}) },
\end{equation}
where $\epsilon_{\text{+},i}$ is the tangential component of the ellipticity of the $i$-th source galaxy after bias correction and $w_i$ is the weight assigned to the source ellipticity. The sum runs over the galaxies included in the annulus at transverse proper distance $R$ from the centre, i.e. the position of the brightest galaxy cluster (BCG). $\Sigma_\text{cr}$ is the critical density for lensing, 
\begin{equation} 
\label{eq_Delta_Sigma_2}
\Sigma_\text{cr}=\frac{c^2}{4\pi G} \frac{D_\text{s}}{D_\text{d} D_\text{ds}}, 
\end{equation}
where $c$ is the speed of light in vacuum, $G$ is the gravitational constant, and $D_\text{d}$, $D_\text{s}$ and $D_\text{ds}$ are the angular diameter distances to the lens, to the source, and from the lens to the source, respectively.  As reference cosmological framework, we assumed the concordance flat $\Lambda$CDM model with total matter density parameter $\Omega_\text{M}=0.3$, baryonic parameter $\Omega_\text{B}=0.05$, Hubble constant $H_0=70~\text{km~s}^{-1}\text{Mpc}^{-1}$, power spectrum amplitude $\sigma_8=0.8$ and initial index $n_\text{s}=1$. When $H_0$ is not specified, $h$ is the Hubble constant in units of $100~\text{km~s}^{-1}\text{Mpc}^{-1}$.

The differential surface density $\Delta\Sigma_{+}$ was measured between 0.1 and $\sim 25.12~\text{Mpc}~h^{-1}$ from the cluster centre in 24 radial circular annuli equally distanced in logarithmic space. The binning is such that there are 10 bins per decade, i.e. 10 bins between 0.1 and $1~\text{Mpc}~h^{-1}$. 

The raw ellipticity components of the sources, $e_{\text{m}, 1}$ and $e_{\text{m}, 2}$, were calibrated and corrected by applying a multiplicative and an additive correction,
\begin{equation}
\label{eq_calibration}
e_{i} = \frac{e_{\text{m}, i} - c_i}{1 + {\bar m}}  \, \hspace{1cm} (i=1,2) \, .
\end{equation}
Each selected source galaxy measurement was individually corrected for the estimated additive bias. The multiplicative bias $m$ mostly depends on the shape measurement technique and was identified from the simulated images\cite{hey+al12,mil+al13}. In each annulus, we considered the average $\bar{m}$, which was evaluated taking into account the weight of the associated shear measurement\cite{vio+al15}, 
\begin{equation}
\label{eq_Delta_Sigma_5}
\bar{m}(R) = \frac{\sum_i w_i \Sigma_{\text{cr},i}^{-2} m_i}{\sum_i w_i \Sigma_{\text{cr},i}^{-2}}.
\end{equation}

We identified the population of background galaxies either with a colour-colour selection $g-r-i$\cite{ogu+al12,cov+al14},
\begin{equation} 
\label{eq_col_1}
(g-r < 0.3) \\ 
\, \text{OR} \\
\, (r - i >  1.3) \\ 
\, \text{OR} \\
\, (r - i > g-r)
\, \text{AND} \\ 
\, z_\text{s} > z_\text{lens} +0.05,
\end{equation}
or with criteria based on the photometric redshifts,
\begin{equation} 
\label{eq_zphot_1}
\texttt{ODDS}>0.8 \,\, \text{AND} \,\, z_\text{2.3\%}  >  z_\text{lens} +0.05 \,\, \text{AND} \,\,   z_\text{max} < z_\text{s} < 1.2,
\end{equation}
where the parameter $\texttt{ODDS}$ quantifies the relative importance of the most likely redshift\cite{hil+al12} and $z_\text{2.3\%}$ is the lower bound of the region including the 95.4\% of the probability density distribution. $z_\text{min}=0.2$ $(0.4)$ for the CFHTLenS (RCSLenS) fields.

The SNR of the WL detection was defined in terms of the weighted differential density $\langle \Delta \Sigma_{+} \rangle_{R_\text{min} <R< R_\text{max}}$ in the corresponding radial range,
\begin{equation}
\label{eq_SNR_1}
\text{SNR}=\frac{\langle \Delta \Sigma_+ \rangle_{R_\text{min} <R< R_\text{max}}}{\delta_+},
\end{equation}
where $\delta_+$ includes statistical uncertainty and cosmic noise added in quadrature. For our analysis of the correlated matter, we considered the signal in the range $10<R~[\text{Mpc}~h^{-1}]<25.1$.

\


\noindent \textbf{Lens modelling.}
\noindent The profile of the differential projected surface density of the lens was modelled as
\begin{equation} 
\Delta \Sigma_\text{tot}=\Delta \Sigma_\text{h}+\Delta \Sigma_\text{e} \pm \Delta \Sigma_\text{LSS} \pm \Delta \Sigma_\text{Stat}.
\end{equation}
The main halo responsible for $\Delta \Sigma_\text{h}$ is a smoothly truncated cuspy density profile\cite{bal+al09},
\begin{equation} 
\rho_\text{BMO} = \frac{\rho_\text{s}}{(r/r_\text{s}) (1 +  r/r_\text{s})^2} \left(\frac{r_\text{t}^2}{r^2 +  r_\text{t}^2} \right)^2,
\end{equation} 
where $r_\text{s}$ is the inner scale length, $\rho_\text{s}$ is the characteristic density and $r_\text{t}$ is the truncation radius.  For our analysis, we set $r_\text{t} = 3\,r_{200} $\cite{og+ha11,cov+al14}, and we expressed $r_\text{s}$ and $\rho_\text{s}$ in terms of mass $M_{200}$ and concentration $c_{200}$. The suffix $200$ ($500$) refers to the region wherein the main halo density is $200$ ($500$) times the cosmological critical density at the cluster redshift. At $R\gtrsim10~\text{Mpc}$, the shear fitting analysis is independent of details about main halo truncation and outskirts. 

The contribution of the local environment surrounding the halo is
\cite{og+ta11,og+ha11}
\begin{equation}
\Delta \Sigma_\text{e} (\theta; M, z) = b_\text{e} \frac { \bar{\rho}_\text{M}(z)}{ (1+z)^3  D_\text{d}^2}  \int \frac{l d l}{2 \pi} J_2(l \theta)P_\text{m}(k_l; z),
\label{eq:gamma_t2}
\end{equation}
where $\bar{\rho}_\text{M}$ is the mean cosmological matter density at the cluster redshift, $ \theta$ is the angular radius, $J_n$ is the Bessel function of $n$-th order, and $k_l \equiv l / [ (1+z) D_\text{d}(z) ]$. $ b_\text{e}$ is the environment bias with respect to the linear dark matter power spectrum $P_\text{m}(k_l; z)$ \cite{sh+to99,tin+al10,bha+al13}. We computed $P_\text{m}$ with semi-analytical approximations\cite{ei+hu99}. 

Large scale structure induces a correlated noise. The cross-correlation between two angular bins $\Delta \theta_i$ and   $\Delta \theta_j$ is\cite{sch+al98b,hoe03}
\begin{equation} 
\label{eq_lss_1}
\langle \Delta \Sigma_\text{LSS}(\Delta \theta_i)  \Delta \Sigma_\text{LSS}(\Delta \theta_j) \rangle = 2 \pi  \Sigma_\text{cr}^2 \int_0^{\infty} P_k(l)g(l, \Delta \theta_i) g(l, \Delta \theta_j) \ dl \ ,
\end{equation}
where $P_k(l)$ is the effective projected power spectrum of lensing. The effects of non-linear evolution on the power spectrum were accounted for with standard methods\cite{smi+al03}. The function $g$ is the filter. In an angular bin $\theta_1 < \Delta \theta< \theta_2$,
\begin{equation} 
g=\frac{1}{\pi(\theta_1^2 -\theta_2^2)l} \left[ \frac{2}{l} \left( J_0(l \theta_2)  -J_0(l \theta_1) \right) +\theta_2 J_1(l \theta_2) -\theta_1 J_1(l \theta_1) \right].
\end{equation}

\


\noindent \textbf{Inference.}
In a predetermined cosmological model, the lensing system is characterized by three parameters, $\textbf{p} = (M_{200},c_{200},b_\text{e})$, which we measured with a standard Bayesian analysis\cite{ser+al15_cM}. The posterior probability density function of the parameters given the data $\{{\Delta\Sigma_{+}} \}$ is
\begin{equation} 
p(\textbf{p}| \{{\Delta\Sigma_{+}} \})  \propto {\cal L}(\textbf{p}) p_\text{prior}(\textbf{p}),
\end{equation} 
where $\cal L$ is the likelihood and $p_\text{prior}$ represents a prior.

The likelihood can be expressed as ${\cal L}\propto \exp (-\chi^2)$, where the $\chi^2$ is written as
\begin{equation}
\chi^2 = \sum_{i,j}  \left[ \Delta \Sigma_{+,i}  - \Delta \Sigma_i (\textbf{p}) \right]^{-1} C_{ij}^{-1}    \left[ \Delta \Sigma_{+,j}  - \Delta \Sigma_j (\textbf{p}) \right];
\end{equation}
the sum extends over the radial annuli and the effective radius $R_i$ of the $i$-th bin is estimated as a shear-weighted radius\cite{ser+al17_psz2lens}; $\Delta\Sigma_{+}(R_i)$ is the measured differential surface density and $ \Delta \Sigma (\textbf{p}) = \Delta \Sigma_\text{h}+\Delta \Sigma_\text{e}$.

$\Delta \Sigma_\text{LSS}$ and $\Delta \Sigma_\text{Stat}$ are treated as uncertainties. The total uncertainty covariance matrix is
\begin{equation}
\label{eq_inf_2}
C= C^\text{stat}+ C^\text{LSS},
\end{equation}
where $C^\text{stat}$ accounts for the uncorrelated statistical uncertainties in the measured shear whereas $C^\text{LSS}_{i,j} = \langle \Delta \Sigma_\text{LSS}(\Delta \theta_i)  \Delta \Sigma_\text{LSS}(\Delta \theta_j) \rangle$ is due to LSS, see equation~(\ref{eq_lss_1}).
 
As mass prior, we considered a uniform probability distribution in the ranges $0.05 \le M_{200}/(10^{14}h^{-1}M_\odot) \le 100$, with the distributions being null otherwise.

For the concentration, we considered a lognormal distribution in the range $1<c_{200}<10$, with median value\cite{men+al14},
\begin{equation}
c_{200}= A \left( \frac{1.34}{1+z}\right)^B  \left( \frac{M_{200}}{8\times 10^{14}h^{-1}M_\odot}\right)^C,
\end{equation}
where $A=3.757$, $B=0.288$, and $C=-0.058$. The scatter of the mass-concentration relation is $0.25$ in natural logarithms.

The prior on the bias is either a Dirac $\delta$ function of the peak height $\nu$, $b_\text{e}=b_\text{h}[\nu(M_{200},z)]$\cite{tin+al10} for the $\Lambda$CDM model or an uniform distribution in the range $0.02<b_\text{e} <200$.

\


\noindent \textbf{WL stacking.}
We stacked the lensing measurements of the PSZ2LenS clusters following a standardized approach\cite{joh+al07,man+al08,cov+al14,ser+al17_psz2lens,ser+al15_bias}. The lensing signals of multiple clusters were combined in physical proper length units. The weight factor is mass-independent and the effective mass and concentration of the stacked clusters is unbiased \cite{oka+al13,ume+al14}. Clusters were centred on the respective BCGs. We fitted a single profile to the stacked signal to determine the ensemble properties\cite{ser+al17_psz2lens}.

\



\begin{table}
\centering
{\footnotesize 
\caption{Systematic error budget on the differential surface density measurements between $R=10$ and $25.1~\text{Mpc}~h^{-1}$. Sources of systematics (col.~1) are taken as uncorrelated.}
\label{tab_systematics_psz2lens}
\begin{tabular}[c]{l  l}
\hline
\noalign{\smallskip}  
Source				&  error [\%]  \\
\hline
\noalign{\smallskip}   
Shear measurements	& $\pm3$\\
Photometric redshifts	&  \\
bias &	$\pm2$      \\
scatter &	$\pm2.5$ \\
Line-of-sight projections	& $\pm1$\\
Contamination	and membership dilution		& $\pm2$ \\
Miscentering			& $\sim 0$ \\
Halo modelling			& $\sim 0$ \\
Intrinsic alignment			&  \\
II & $\sim 0$ \\
GI & $-0.6$ \\ 
Total					& $\pm5$ \\
   	 \hline
	\noalign{\smallskip}      
	\end{tabular}
	}
\end{table}

\noindent \textbf{WL systematics.}  
Systematic uncertainties on the shear signal are listed in table~\ref{tab_systematics_psz2lens}. Errors not accounted for in equation~(\ref{eq_inf_2}) can be quantified by an analysis of the full PSZ2LenS sample\cite{ser+al17_psz2lens}. The main contributors to the error budget are the calibration uncertainty of the multiplicative shear bias, the photo-$z$ accuracy and precision, and the selection of the source galaxies. 

The multiplicative bias is well controlled, but a calibration uncertainty in the shape measurements can persist at the level of a few per cents. By detailed comparison of separate shape catalogues\cite{jar+al16}, the systematic uncertainty can be estimated in $\delta m \sim 0.03$. 

Cluster members or foreground galaxies can dilute the lensing signal. Our conservative selection criteria based on either photometric redshifts or colour-colour cuts suffer by a $\lesssim 2\%$ contamination.

Miscentring can underestimate the shear signal at small scales and affect the concentration measurement\cite{joh+al07}. The effect is however negligible at large scales $\gtrsim 10~\text{Mpc}$.

Photometric redshift systematics can impact weak lensing analyses by biasing the estimation of the surface critical density. As source redshifts, we considered the peak of the probability density, as applicable to well behaved and single peaked distributions. The systematic error associated to either a bias or a scatter in the photo-$z$ estimates was estimated for PSZ2LenS clusters with simulations\cite{ser+al17_psz2lens}.

Improper halo modelling can affect the mass and concentration estimate at a few percents\cite{ser+al16_einasto}. However, as far as the truncation of the main halo is modelled, the effect on $b_\text{e}$ is negligible.

The role of cluster projection is marginal too. Two clusters that fall along the same line of sight may be blended by the SZ cluster finder into a single, larger cluster. Whereas the Compton parameters add approximately linearly, the lensing amplitude $\Delta \Sigma_+$ is a differential measurement and the estimated $b_\text{e}$ of the blended system can be well below the sum of the two the aligned halos. However, the chance to have two or more  {\it Planck} clusters aligned is $\sim 5\%$. At $z=0.616$, the systematic error is then negligible ($< 1\%$).

Intrinsic galaxy alignment of physically nearby galaxies (II) can contaminate the signal. Furthermore, background galaxies experience a shear caused by the foreground tidal gravitational field. If the foreground galaxy has an intrinsic ellipticity that is linearly correlated with this field, shape and shear are correlated (GI). In the intrinsic alignment model\cite{br+ki07,hey+al13}, the power spectra of intrinsic alignment II and GI  are proportional to the matter power spectrum,
\begin{equation}
P_\text{II} = F_z^2 P_\delta \ , \ \ P_\text{GI} = F_z^2 P_\delta \ , 
\end{equation}
with 
\begin{equation}
F_z = - C_1 \rho_\text{Cr}(z=0) \frac{\Omega_\text{M}}{D(z)} , 
\end{equation}
where 
$D(z)$ is the linear growth factor normalized to unity today and $C_1 = 5 \times 10^{-14} h^{-2}M_\odot^{-1}\text{Mpc}^3$. At $z=0.616$, the combined systematic error from II and GI is then negligible ($< 1\%$).

The total level of systematic uncertainty is $\sim5\%$. 

\


\begin{figure}
\resizebox{\hsize}{!}{\includegraphics{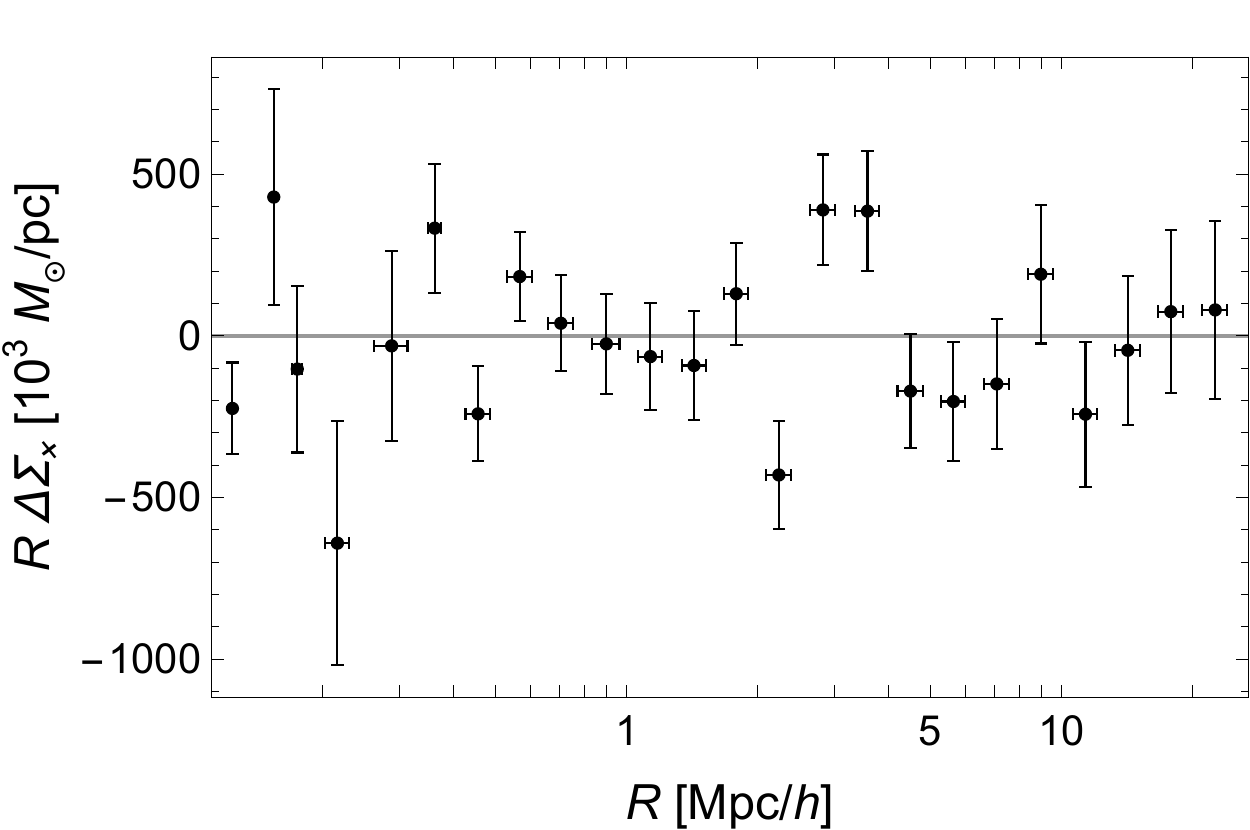}}
\caption{\textbf{Cross-component of the shear profile.}  Shown is the renormalised cross-component of the differential surface density $\Delta \Sigma_\times$ of PSZ2~G099.86+58.45. Error bars are as in Fig.~\ref{fig_PSZ0478_Delta_Sigma}.
}
\label{fig_PSZ0478_Delta_Sigma_X}
\end{figure}

\begin{table}
\caption{Here reported the WL measurements of PSZ2~G099.86+58.45 and the results of the inference analysis under different assumptions (as detailed in col.~1). We list the differential surface density $\Delta\Sigma_+$ between $R=10$ and $25.1~\text{Mpc}~h^{-1}$, in units of $M_\odot~h~\text{pc}^{-2}$ in col.~2, the SNR in col.~3 and the estimated environment bias in col.~4.}
\label{tab_systematics_0478}
\centering
{\footnotesize
\begin{tabular}[c]{l  r@{$\,\pm\,$}l   r  r@{$\,\pm\,$}l }
\hline
	\noalign{\smallskip}  
   &  \multicolumn{2}{c}{$\Delta\Sigma_+$}	& SNR	&  \multicolumn{2}{c}{$b_\text{e}$}    \\
    	 \hline
	\noalign{\smallskip}      
reference                 	&	32.         	&	9.                	&	3.5	&	78. 	&	11.  	\\
\noalign{\smallskip}   
\multicolumn{6}{l}{Shape bias} \\
\noalign{\smallskip}   
WL pass                   	&	24.         	&	10.               	&	2.4	&	58. 	&	14.  	\\
$c_{2}=0$                 	&	32.         	&	9.                	&	3.5	&	78. 	&	11.  	\\
\noalign{\smallskip}   
\multicolumn{6}{l}{Quadrants} \\
\noalign{\smallskip}   
NW                        	&	55.         	&	12.               	&	4.5	&	117.	&	22.  	\\
NE                        	&	3.         	&	12.               	&	0.3	&	20. 	&	14.  	\\
SE                        	&	36.         	&	12.               	&	3.1	&	80. 	&	20.  	\\
SW                        	&	33.         	&	12.               	&	2.7	&	87. 	&	22.  	\\
\noalign{\smallskip}   
\multicolumn{6}{l}{Background selection} \\
\noalign{\smallskip}   
$gri$                     	&	31.         	&	9.                	&	3.3	&	73. 	&	11.  	\\
$z_\text{phot}$           	&	31.         	&	12.               	&	2.5	&	71. 	&	22.  	\\
\noalign{\smallskip}   
\multicolumn{6}{l}{photo-$z$ uncertainties} \\
\noalign{\smallskip}   
bias  	&	32.         	&	11.               	&	2.9	&	77. 	&	15.  	\\
scatter	&	38.         	&	11.               	&	3.4	&	82. 	&	16.  	\\
\noalign{\smallskip}   
\multicolumn{6}{l}{Centering and member dilution} \\
\noalign{\smallskip}   
SZ centred                	&	32.         	&	9.                	&	3.4	&	81. 	&	11.  	\\
$R>0.5~\text{Mpc}~h^{-1}$      	&	32.         	&	9.                	&	3.5	&	77. 	&	11.  	\\
\noalign{\smallskip}   
\multicolumn{6}{l}{Cosmology} \\
\noalign{\smallskip}   
WMAP9                     	&	32.         	&	9.                	&	3.5	&	73. 	&	10.  	\\
NL-$P_\delta$             	&	32.         	&	9.                	&	3.5	&	71. 	&	9.   	\\
	\hline
	\end{tabular}
}
\end{table}

\noindent \textbf{WL stress tests.} 
We checked for potential residual sources of erros in the WL analysis of PSZ2~G099.86+58.45. Not properly corrected systematics can affect the cross component of the shear signal. We verified that it is consistent with zero as expected, see Fig.~\ref{fig_PSZ0478_Delta_Sigma_X}. The $p$-value of the null hypothesis is 0.08.

We then repeated the analysis of the tangential component under a series of assumptions to check whether the systematic level is sufficiently smaller than the statistical noise, see table~\ref{tab_systematics_0478}.

The cluster catalogue and the shape measurements in our analysis are taken from completely different data sets. The distribution of lenses is then uncorrelated with residual systematics in the shape measurements\cite{miy+al15}. However, significant residual errors can hamper the shape measurements if the PSF (point spread function) is very anisotropic. We then considered only fields that passed the CFHTLenS weak lensing selection (WL pass), i.e. fields with a low level of PSF anisotropy contamination as estimated from the analysis of the star-galaxy cross correlation\cite{hey+al12,erb+al13}. The result is in very good agreement with the reference case, as also confirmed by the analysis performed disregarding the additive bias correction, i.e. by putting $c_2=0$. 

To check if the excess signal can be associated to a single structure nearby or in projection, we measured the shear signal in sectors. The signal in the North-West (NW), North-East (NE), South-East (SE) and South-West (SW) quadrants is compatible with the signal in the full field of view, whereas the bias exceeds the average $\Lambda$CDM prediction in each sector. This confirms that the excess is not related to foreground or background massive halos.

The selection of background galaxies was checked by comparing results obtained using either the color-color, see equation~(\ref{eq_col_1}), or the photo-$z$ method, see equation~(\ref{eq_zphot_1}). 

The effects of centering or cluster member dilution were checked by considering the SZ centroid as lens center or excising the inner region at $R <0.5~\text{Mpc}~h^{-1}$.

The extent of errors affecting the photo-$z$ estimates was checked either adding a positive bias $+0.01(1+z)$ to the peak of the redshift distributions of the galaxies in the field or randomly scattering the peaks with a Gaussian distribution with standard deviation $\sigma_z=0.04 (1+z)$.

Variations in the shear signal due to the cosmological model are negligible too, as checked by considering the cosmological parameters  from the nine-year Wilkinson Microwave Anisotropy Probe (WMAP9) observations\cite{hin+al13}. 

Halo bias is usually defined with respect to the liner power spectrum. However, even considering non-linearities, the estimated bias is still much larger than the average $\Lambda$CDM prediction (NL-$P_\delta$).

\


\begin{figure}
\resizebox{\hsize}{!}{\includegraphics{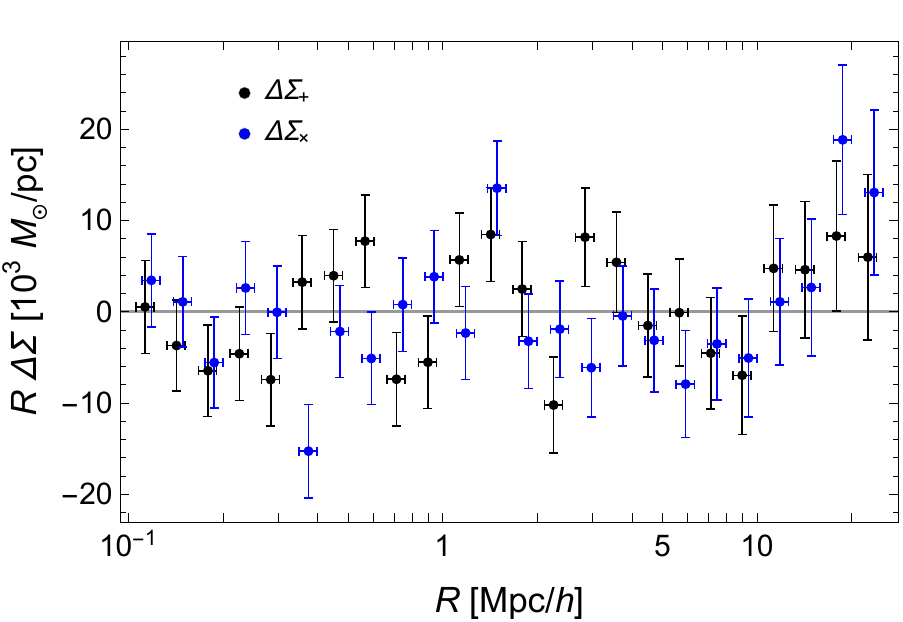}}
\caption{\textbf{Lensing profile around random points.}  Shown are the renormalised tangential (black, $\Delta \Sigma_+$) and cross (blue, $\Delta \Sigma_\times$) component of the differential surface density of the stacked signal collected around 1000 random locations in the CFHT-W3 field at $z=0.616$. Error bars are as in Fig.~\ref{fig_PSZ0478_Delta_Sigma}.  Blue points are horizontally shifted for visualization purposes.
}
\label{fig_PSZ0478_Delta_Sigma_random}
\end{figure}

\begin{figure}
\resizebox{\hsize}{!}{\includegraphics{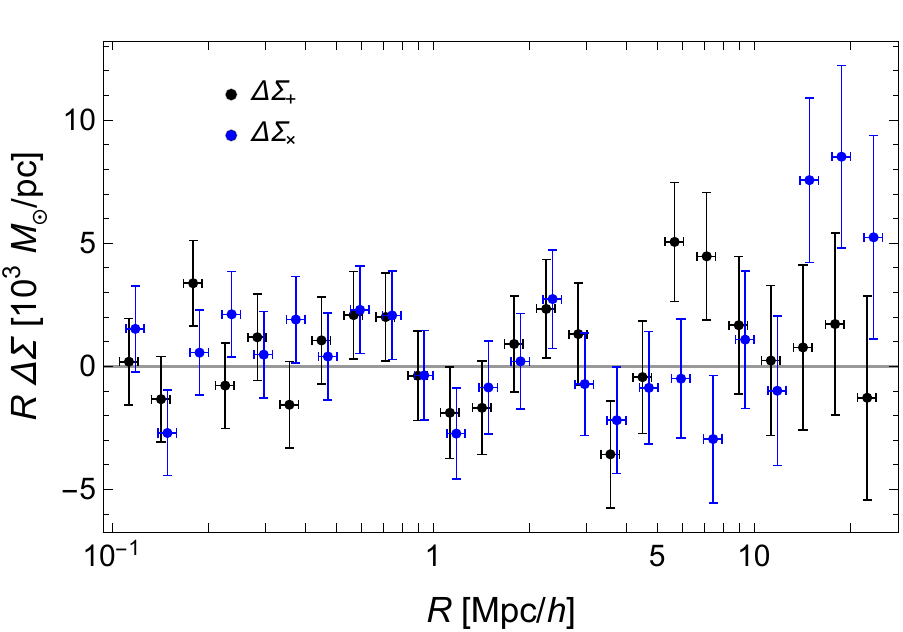}}
\caption{\textbf{Stacked lensing profile of PSZ2LenS-like random catalogs.}  Shown are the renormalised tangential (black, $\Delta \Sigma_+$) and cross (blue, $\Delta \Sigma_\times$) component of the differential surface density of the stacked signal collected around 100 random PSZ2LenS-like catalogs. Error bars are as in Fig.~\ref{fig_PSZ0478_Delta_Sigma}. Blue points are horizontally shifted for visualization purposes.
}
\label{fig_PSZ2LenS_Delta_Sigma_random}
\end{figure}

\noindent \textbf{Random pointings.} 
General features of the large scale structure can be studied by extracting the signal around random points with the same procedure used for the cluster analysis. We measured the differential density associated to 1000 random positions in the CFHT-W3 field at redshift $z=0.616$ and we stacked the signals. Both the tangential and the cross-component of the shear are consistent with a null signal, see Fig.~\ref{fig_PSZ0478_Delta_Sigma_random}. The $p$-value of the null hypothesis is 0.30 (0.12) for the tangential (cross) component. This further confirms that the main systematics have been eliminated and that the signal excess around PSZ2~G099.86+58.45 is significant, see Fig.~\ref{fig_PSZ0478_histo_DeltaSigma_sim}. 

Residual systematics can affect the stacking analysis due to incomplete annuli for clusters near the border of the field of view or to partially overlapping regions for nearby clusters. We computed the stacked signal for 100 PSZ2LenS-like catalogs of 35 random sources reproducing the input redshift distribution and the field locations. Both the tangential and the cross component of the stacked shear are very well consistent with a null signal, see Fig.~\ref{fig_PSZ2LenS_Delta_Sigma_random}. The $p$-value of the null hypothesis is 0.49 (0.25) for the tangential (cross) component.


\


\begin{figure}
\resizebox{\hsize}{!}{\includegraphics{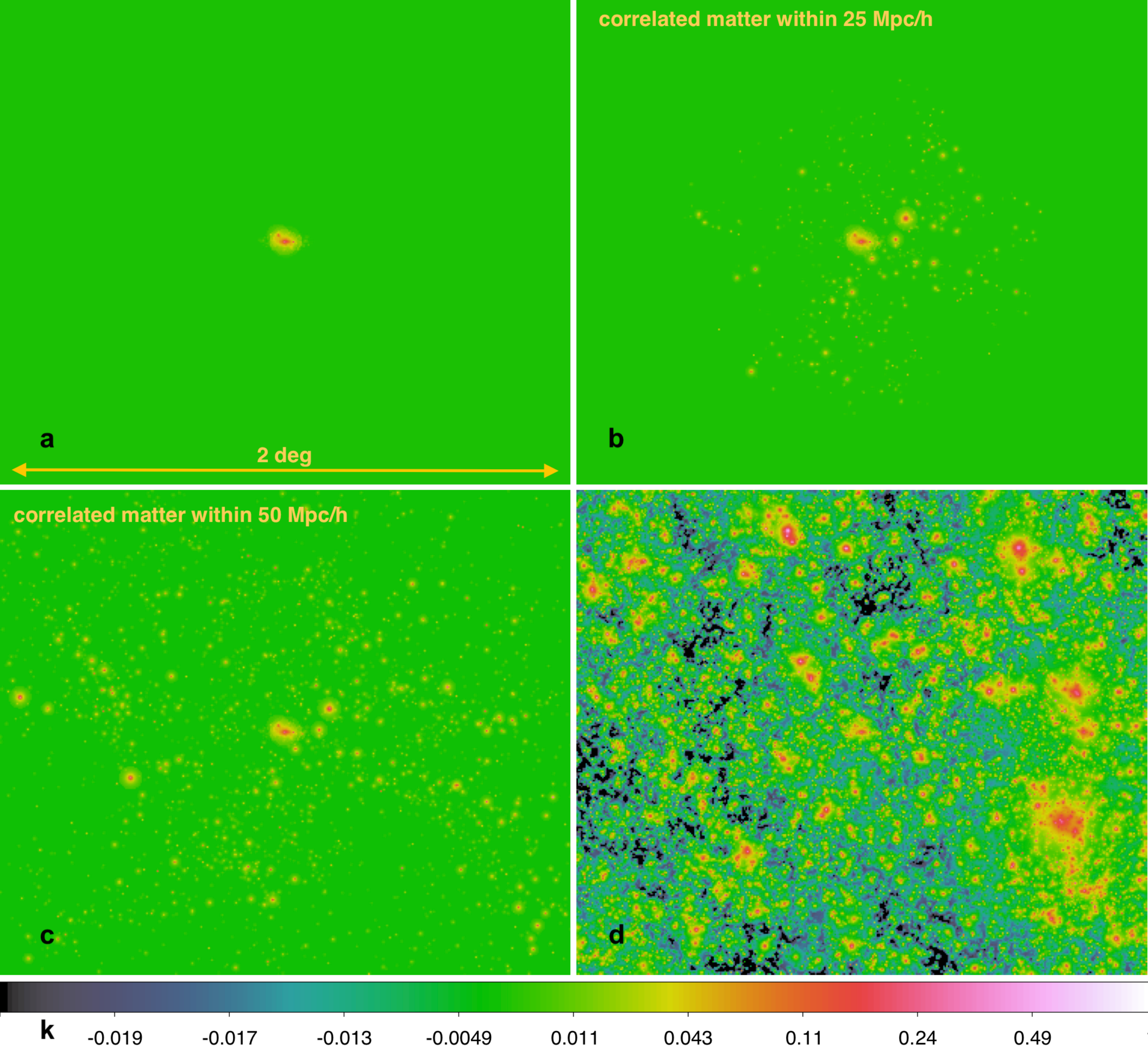}}
\caption{\textbf{Simulated map.}  Shown is a simulated convergence map. The upper left panel shows an isolated cluster with a mass of $\sim 7.4\times10^{14} M_\odot h^{-1}$ at $z\sim 0.67$. Upper right and bottom left panels: central halo plus matter within a comoving distance of 25 or $50~\text{Mpc}~h^{-1}$, respectively. Bottom right panel: all the correlated and uncorrelated structures present in the field of view of the cluster.
}
\label{fig_map_sim}
\end{figure}

\begin{figure}
\resizebox{\hsize}{!}{\includegraphics{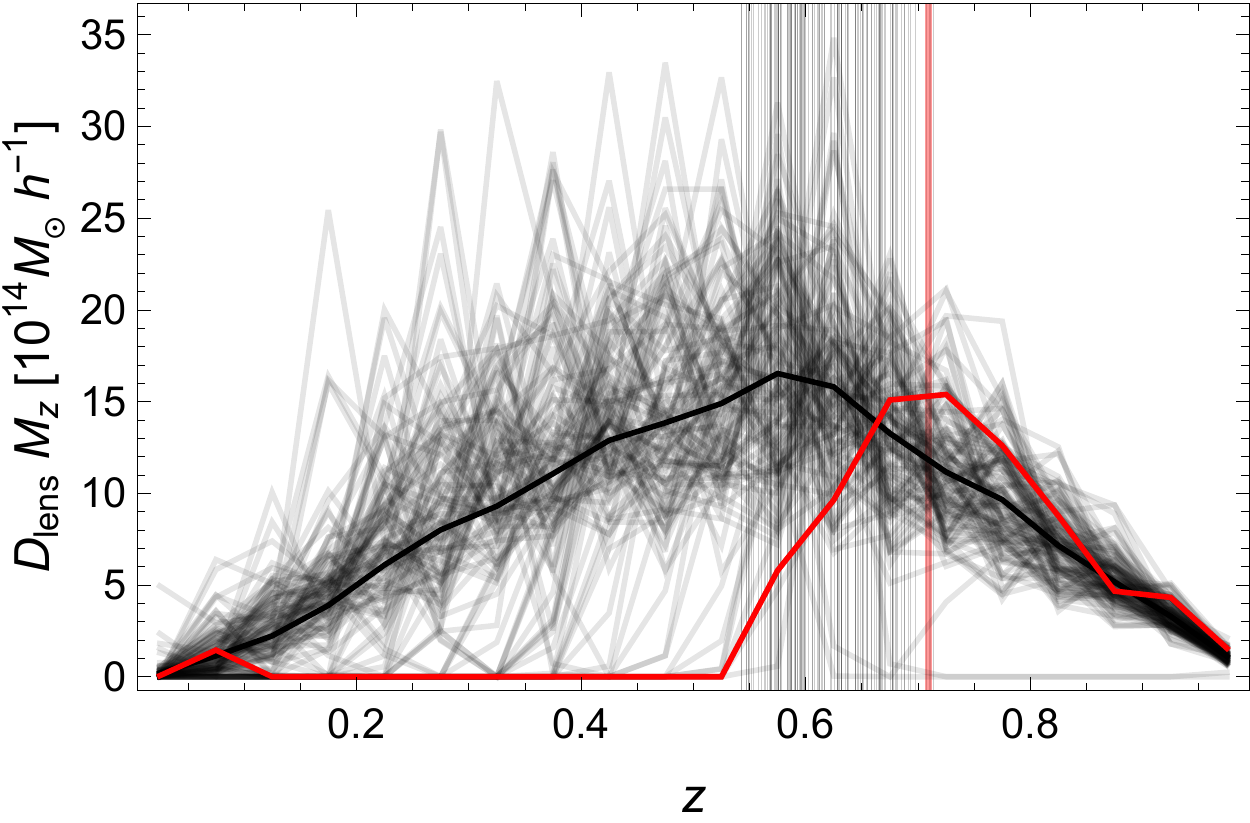}}
\caption{\textbf{Simulated mass distributions}. Shown is the total matter in simulated collapsed halos as a function of redshift. We added the masses of halos more massive than $7\times10^{11}M_\odot/h$ in redshift slices  of thickness $\Delta z=0.05$ and within an aperture of 1~deg. The matter is normalized by the lensing kernel $D_\text{lens}=D_\text{ds}/D_\text{s}$. The vertical lines mark the redshift of the simulated lens. The black, grey and red lines correspond to the simulation average, the individual systems and the system with the highest shear signal between 10 and $25.1~\text{Mpc}~h^{-1}$, respectively.
}
\label{fig_PSZ0478_Mz_sim}
\end{figure}

\noindent \textbf{Numerical simulations of the WL signal.} 
We generated the WL signal of cluster and large scale structure using a sample of self-consistent halo model simulations. We produced a set of halo catalogues within past-light cone simulations up to $z=1$ using \texttt{Pinocchio}\cite{mon+al13}. \texttt{Pinocchio} is a fast code to generate catalogues of cosmological dark matter halos starting from an initial power spectrum and perturbing it using the Lagrangian perturbation theory (LPT) model. For this work we performed the \texttt{Pinocchio} simulations using the 3LPT approximation\cite{mun+al17}. The large scale matter density distribution in halos by \texttt{Pinocchio} accurately reproduces the results of $N$-body simulations. 

Out of 512 light-cone realisations, we extracted a sample of 128 halos with mass and redshift similar to PSZ2~G099.86+58.45. We constructed the effective convergence map of the full light-cone of the main cluster plus correlated and uncorrelated systems up to redshift $z=1$ (Fig.~\ref{fig_map_sim}), using the \texttt{MOKA}\cite{gio+al12a} and the \texttt{WL-MOKA}\cite{gio+al17} tools. In particular, we constructed the effective convergence map of the cluster using \texttt{MOKA}, which generates triaxial systems populated with dark matter substructures mimicking halos from numerical simulations. We located a BCG, which was modelled using a Jaffe profile, in the centre of the cluster. The halo dark matter distribution is adiabatically contracted as consequence. Correlated matter and uncorrelated LSS are modelled as isolated NFW\cite[Navarro, Frenk \& White,][]{nfw96} halos with a mass-concentration relation consistent with field halos\cite{men+al14}. The aperture of our field of view is 3 degrees by side; by construction our light-cones are pyramids where the observer is located at the vertex and the base is at a fixed source redshift, $z_\text{s}=1$. We computed the shear field from the convergence maps using fast Fourier methods. 

We finally measured the reduced shear profile around the cluster centre assuming a source density of 32 galaxies per arcminute$^2$. The very large background density of sources make it sure that the signal measured in the simulated systems is due to real features in the matter distribution. It is not due to measurement uncertainties.

Alternatively to the shear signal, the cluster environment was studied by analyzing the mass distribution as a function of redshift. We measured the total matter (except for the central cluster) collapsed in halos above the minimum threshold of $7\times10^{11}M_\odot/h$. The redshift slices were $\Delta z= 0.05$ thick and we considered only halos in limited angular radial apertures ranging from 0.15 to 1.5 degree. In Fig.~\ref{fig_PSZ0478_Mz_sim}, we plot the result for an aperture of 1 degree. The mass was rescaled by the lensing distance kernel $D_\text{lens}=D_\text{ds}/D_{s}$.


\

\noindent \textbf{X-ray analysis.} 
The X-ray analysis was performed on archival {\it XMM}-Newton data observed on November 8th, 2013. We applied the standard calibration in order to obtain the event lists for the EPIC detector, using the \texttt{cifbuild}, \texttt{odfingest}, and \texttt{emchain} packages \cite{sno+al08}. Background sources were excluded with the \texttt{cheese} tool. We preliminarily applied a standard filtering with the \texttt{mos-filter} and \texttt{pn-filter} package for the MOS and PN detector, respectively, in order to check the contamination by soft-proton background. The high number of CCDs (three) in the anomalous mode for the MOS1 detector led us to consider only the MOS2 and the PN detectors for our analysis. The particle background model for our starting images and spectra was produced with the \texttt{mos\_back} and \texttt{pn\_back} packages.

We then selected the time intervals less contaminated by soft-proton background. Using images in the soft and hard {\it XMM} bands, we identified and removed extra X-ray sources located in the region of interest. Finally, we obtained the spectral files for the source, the background, and the instrumental responses. 

The spectral analysis was performed with \texttt{XSPEC}\cite{anr96}. We considered an absorbed \texttt{APEC} thermal model for the cluster component, with metal abundance fixed at $Z=0.3 Z_{\odot}$. 

We took into account different background sources: i) an unabsorbed thermal component representing the local hot bubble\cite{ku+sn08}; ii) an absorbed thermal component which models the intergalactic medium and the cool halo\cite{ku+sn08}; and iii) an absorbed power-law with spectral index $\alpha = 1.46$ representing the unresolved background of cosmological sources\cite{tak+al11}. In addition, we included emission lines rising from the solar wind charge exchange at 0.56 and 0.65 keV. We finally included three Gaussian models in order to consider bright fluorescent lines at 1.49, 1.75 and 8~keV, due to the K$\alpha$ of the Al, Si and Cu, respectively. 

The spectra were finally fitted in the range [0.4--7.2]~keV. The X-ray temperature was converted in mass exploiting calibrated scaling relations\cite{vik+al09}.


\

\noindent \textbf{Optical spectroscopy.} 
The spectroscopic redshift analysis was performed under an International Time Project (ITP13-08) from August 2012 to July 2013 \cite{planck_int_XXXVI}. We preliminarily calculated the photometric redshift of the cluster using archival Sloan Digital Sky Survey (SDSS) DR12 data \cite{ala+al15}. We identified likely cluster members showing coherent colours in agreement with $z_\text{phot}=0.63\pm 0.03$.

In order to confirm the cluster and obtain an estimate of the galaxy velocity dispersion, we performed spectroscopic observations using the OSIRIS spectrograph of the 10.4m GTC telescope, at Roque de los Muchachos Observatory (ORM) in Canary Island, during March 2014. We obtained spectroscopic redshift for 8 galaxy members by setting the long-slit in two position angles. The exposure time was $3~\text{ks}$ for each position. The full wavelength range, 4000-9000~\AA, was covered with a resolution $R\sim500$. 

The spectroscopic data reduction was performed using standard  \texttt{IRAF} tasks\cite{tod86}; radial velocities were obtained using \texttt{XCSAO}, i.e. the cross-correlation technique \cite{ton79} implemented in the \texttt{IRAF} task \texttt{RVSAO}, with six spectrum templates of different galaxy morphologies: E, S0, Sa, Sb, Sc, and Irr \cite{ken92}.

We measured radial velocities for 8 galaxy cluster members, including the BCG at $\text{RA}=213.696611\deg$, $\text{DEC}=54.784321\deg$ (J2000) and $z_\text{BCG}=0.6139\pm0.0002$. In addition, we also considered 4 spectroscopic redshifts from SDSS DR12. All galaxy members are placed within $2.5~\text{Mpc}$ from the cluster centre and show velocities within $\pm 2500~\text{km~s}^{-1}$ from the BCG. The full spectroscopic dataset reveals that PSZ2~G099.86+58.45 is at $z_\text{spec}=0.616\pm0.001$.

The galaxy velocity dispersion $\sigma_\text{v}=680^{+160}_{-130}~\text{km~s}^{-1}$ was estimated using the gapper scale estimator \cite{bee+al90}. This estimate can be used as a mass proxy through a calibrated scaling relation \cite{evr+al08}.

\

\noindent \textbf{Data availability.} 
The data that support the findings of this study are available at  \url{http://pico.oabo.inaf.it/~sereno/} or from the corresponding author upon reasonable request. The WL data were obtained with MegaPrime/MegaCam, a joint project of CFHT and CEA/IRFU, at the Canada-France-Hawaii Telescope (CFHT) which is operated by the National Research Council (NRC) of Canada, the Institut National des Sciences de l'Univers of the Centre National de la Recherche Scientifique (CNRS) of France, and the University of Hawaii. 
The CFHTLenS and RCSLenS catalogues including photometry and lensing shape information are publicly available at \url{http://www.cadc-ccda.hia-iha.nrc-cnrc.gc.ca/en/community/CFHTLens/query.html} and \url{http://www.cadc-ccda.hia-iha.nrc-cnrc.gc.ca/en/community/rcslens/query.html}, respectively. 

{\footnotesize
\renewcommand{\refname}{\small References}

}

\section*{\small Acknowledgements}
\footnotesize
The authors thank Jos\'e Alberto Rubi\~no Mart\'in for coordinating the spectroscopic campaign and Luca D'Avino for suggestions on the rendering of Fig.~\ref{fig_PSZ0478_map}. SE and MS acknowledge financial support from the contracts ASI-INAF I/009/10/0, NARO15 ASI-INAF I/037/12/0, ASI 2015-046-R.0, and ASI-INAF n.2017-14-H.0. CG acknowledges support from the Italian Ministry for Education, University, and Research (MIUR) through the SIR individual grant SIMCODE, project number RBSI14P4IH, and the Italian Ministry of Foreign affairs and International Cooperation, Directorate General for Country Promotion for Country Promotion. LI acknowledges support from the Spanish research project AYA 2014-58381-P. LM thanks the support from the grants ASI n.I/023/12/0 and PRIN MIUR 2015.  AF, AS, RB  acknowledge financial support from the Spanish Ministry of Economy and Competitiveness (MINECO) under the AYA2014-60438-P, the ESP2013-48362-C2-1-P and the 2011 Severo Ochoa Program MINECO SEV-2011-0187 projects. This article includes observations made with the Gran Telescopio Canarias (GTC) operated by Instituto de Astrof\'isica de Canarias (IAC) with telescope time awarded by the CCI International Time Programme at the Canary Islands observatories (program ITP13-8). The simulations were run on the Marconi supercomputer at Cineca thanks to the projects IsC10\_MOKAlen3 and IsC49\_ClBra01. 

\section*{\small Author contributions}
\footnotesize
\noindent 
All authors contributed to the interpretation and presentation of the results. MS: lead author; project concept, planning, and design; writing; lensing, statistical, and cosmological analyses. CG: numerical simulations. LI: X-ray analysis. FM, AV: cosmological analysis. SE, LM: planning and interpretation. GC: cluster sample selection. AF, AS, RB: galaxy kinematics.

\section*{\small Additional information}
\footnotesize
\noindent {\bf Correspondence and requests for materials} should be addressed to MS at \href{mailto:mauro.sereno@oabo.inaf.it}{mauro.sereno@oabo.inaf.it}. \\

\section*{\small Competing interests}
\footnotesize
\noindent The authors declare no competing financial interests. \\

\end{document}